\documentclass[preprint,12pt]{elsarticle}
\usepackage{amsmath}
\usepackage{amssymb}
\usepackage[pdftex,colorlinks]{hyperref}
\usepackage{multirow}

\biboptions{compress}

\begin{document}

\begin{frontmatter}

\title{A note on quantum McEliece public-key cryptosystem}

\author[label1]{Li Yang\corref{1}}\ead{yang@is.ac.cn}
\cortext[1]{Corresponding author.}
\address[label1]{State Key Laboratory of Information Security, Institute of Information Engineering, Chinese Academy of Sciences,
Beijing 100093, China}
\author[label2]{Min Liang}
\address[label2]{Data Communication Science and Technology Research Institute, Beijing 100191, China}

\begin{abstract}
Inspired by Fujita's analysis [Quantum inf. \& comput. 12(3\&4), 2012], we suggest a twice-encryption scheme to improve the security of the original quantum McEliece public-key encryption algorithm.
\end{abstract}


\end{frontmatter}


The security of McEliece public-key cryptosystem (PKC) ~\cite{mceliece1978} relies on the difficulty of decoding general linear codes, an NP-complete problem. McEliece PKC has been analyzed for more than thirty years, and is still regarded as a secure one. It is generally accepted that quantum computer cannot solve NP-complete problems in polynomial time, so McEliece PKC is expected to be one that can resist quantum attacks.  Ref.~\cite{yang2003} extends it to a public-key encryption algorithm for quantum messages. This quantum PKC degenerates to original McEliece PKC while the plaintext state is restricted to single superposition component, then it is secure to encrypt classical messages. Based on the idea of this quantum PKC, we introduce a new concept named induced trapdoor one-way quantum transformation~\cite{yang2011} and a general framework of quantum PKC.

Recently, Fujita \cite{fujita2012} proposes another quantum PKC based on stabilizer codes, and finds a vulnerability of the original quantum McEliece PKC in Ref.~\cite{yang2003}. His argument is that since the attacking with only cipher state $\sum_m\alpha_m|mG\oplus e\rangle$ can be reduced to attacking with the quantum state
\begin{equation}
X(eG^{-})\sum_m\alpha_m|m\rangle,
\end{equation}
and the statistical distribution of $X(eG^{-})\sum_m\alpha_m|m\rangle$ over conjugate bases is the same as that of the quantum message $\sum_m\alpha_m|m\rangle$, the adversary can get some information of $\sum_m\alpha_m|m\rangle$ from $X(eG^{-})\sum_m\alpha_m|m\rangle$, though these two sets of amplitudes are not the same.
\\

The vulnerability of PKC in Ref.~\cite{yang2003} is due to the lack of phase encryption. We suggest here a twice-encryption scheme to overcome this problem. Alice firstly uses one public key $(G,t)$ to encrypt a $k$-qubit message $\sum_m\alpha_m|m\rangle$, and obtains an $n$-qubit state $\sum_m\alpha_m|mG\oplus e\rangle$, then performs Hadamard transformation $H^{\otimes n}$, and obtains an $n$-qubit state, finally she encrypts the $n$-qubit state with another public key $(G_2,t_2)$.

Now we consider its security. Let $G_{1}^{-}$,$G_{2}^{-}$are right inverse matrices of $G_{1}$and$G_{2}$, respectively. Similar to the analysis presented in \cite{fujita2012}, we can find that the cipher state only attack can be realized via attacking the state
\begin{equation}
X(e_2G_2^{-})H^{\otimes n}\sum_m\alpha_m|mG\oplus e\rangle.
\end{equation}
From the relations
\begin{eqnarray}
X(e_2G_2^{-})H^{\otimes n}\sum_m\alpha_m|mG\oplus e\rangle = H^{\otimes n}Z(e_2G_2^{-})\sum_m\alpha_m|mG\oplus e\rangle \nonumber\\
= H^{\otimes n}\sum_m\alpha_m(-1)^{(e_2G_2^{-})\cdot(mG\oplus e)}|mG\oplus e\rangle,
\end{eqnarray}
we know that the attacker can perform $H^{\otimes n}$ on the above quantum state, and then perform a transformation defined by $G^{-}$, and finally obtain the state $X(eG^{-})\sum_m\alpha_m(-1)^{(e_2G_2^{-})\cdot(mG\oplus e)}|m \rangle$.
 We can see that both bit-flip errors and phase errors are introduced into the final state, the attacker can only obtain a state of probability distribution identical with that of the cipherstate of the first encryption. Thus, the twice-encryption scheme can overcome the vulnerability of PKC in Ref.~\cite{yang2003}. This is similar to the private quantum channel\cite{ambainis}, in which $2$-bit key is used to encrypt each qubit perfectly \cite{boykin}.
In the same way, the quantum PKCs in Ref.~\cite{yang2011} can also be improved with twice-encryption method. It can be seen that the twice-encryption scheme is still simpler than that suggested in Ref.~\cite{fujita2012}.
\\

Some points in Ref.~\cite{fujita2012} are worth to be clarified.

\begin{itemize}
  \item {\it PKC in Ref.~\cite{yang2003} is insecure while encrypting classical messages.}
\end{itemize}

Ref.~\cite{fujita2012} argues that the PKC in Ref.~\cite{yang2003} is insecure while encrypting classical messages. We can see that it is suffice to consider the state $X(eG^{-})|m\rangle$
in this case.
Because $X(eG^{-})|m\rangle=|m\oplus eG^{-}\rangle$,
attacking cipher state $X(eG^{-})|m\rangle$ is equivalent to attacking $m\oplus eG^{-}$.
Decoding $m\oplus eG^{-}$ is at least the same hard as decrypting $mG\oplus e$
which is just the cipher of classical McEliece PKC.
Now,we prove the fact that decoding $m\oplus eG^{-}$ is as hard as solving a ``learning parity with noise" (LPN) problem, which is NP-complete.

Here $G=SG_0P$, where $S,P$ are both invertible matrices, $G_0$ is generator matrix of Goppa code and is full row rank, so $G$ is also full row rank, and then it has Moore-Penrose inverse. Suppose $G_1^-$ is one of Moore-Penrose inverses of $G$ satisfying $GG_1^-=I$. In fact, $G_1^-$ can be obtained by solving the linear equations $GX=I$. Then all the Moore-Penrose inverses of $G$ can be written as the form
$$G^-=G_1^-\oplus U\oplus G_1^-GU,$$
where $U$ is any $n\times k$ binary matrix. It can be verified that $GG^-=I$.
In classical McEliece PKE scheme, the cipher $c$ and plaintext $m$ satisfy the relation $c=mG\oplus e$, where $e$ is a binary row vector of weight $t$.
Suppose Eve finds another Moore-Penrose inverse of $G$, denoted as $G_2^-$, then he can compute $cG_2^-=m\oplus eG_2^-$.
Denote $G_2^-=(e_1\cdots e_k)$, where each $e_i$ is a binary column vector. Then $cG_2^-$ can be represented as $(m_1\oplus e\cdot e_1),\cdots,(m_k\oplus e\cdot e_k)$.
If each column $e_i$ of $G_2^-$ has more zeros (it means the Hamming weight of $e_i$ is small enough), $e\cdot e_i$ would equal to $0$ with large probability, then its $i-$th bit $m_i\oplus e\cdot e_i$ would reveal the $i-$th bit of original plaintext with large probability.
Notice that $e_i=g_i\oplus (I\oplus G_1^-G)u_i$, where $g_i$ and $u_i$ are the $i-$th column of $G_1^-$ and $U$ separately.
Here $g_i$ and $I\oplus G_1^-G$ are known, but $u_i$ is unknown. Now Eve have to face a problem: finding $u_i$,
such that $g_i\oplus (I\oplus G_1^-G)u_i$ has weight smaller than a given value. This is just a LPN problem,
which is a NP-complete problem. This problem can be seen from another view. $I\oplus G_1^-G$ is a $n\times n$ matrix, and $g_i,u_i$ are two $n\times 1$ vectors, but $u_i$ is unknown. So the above problem can be restated as follow: how to select some columns of $I\oplus G_1^-G$, such that their summation is closest to vector $g_i$? This is just a closest vector problem (CVP), which is a NP-hard problem.
Thus, attacking via $m\oplus eG^{-}$ is as hard as solving an NP-complete problem. NP-complete problem is believed to be intractable by quantum computer.
Then, we know that the PKC in Ref.~\cite{yang2003} is secure while encrypting classical messages.

We have tried numerical experiment following the above attack, and it seems that this kind of attack is invalid.
This attack is reduced to an optimal problem (CVP): finding $u_i$, such as $g_i\oplus (I\oplus G_1^-G)u_i$ (notice that it equals $e_i$) has weight smaller than a given value.
Suppose the parameters $n=1024,k=524,t=50$ in the McEliece PKC scheme.
Firstly, because $u_i$ has $2^{1024}$ choices, both the exhaustive search and random search are not realistic. While choosing some small parameters such as $n=60,k=30$, the exhaustive search can reduce the weight of $e_i$ to $1$ with probability $2\%$, and the random search may be slightly better.  With the greedy search, we obtain $e_i$ of weight $225$ on average. In this case, $Pr[e\cdot e_i=0]\approx 0.5+0.1\times 10^{-13}$, here $e$ is a $n$-bit random vector of weight $t=50$. Thus, the attack presented here is invalid.
$\square$

\begin{itemize}
  \item {\it The state $X(eG^{-})\sum_m\alpha_m|m\rangle$ is equivalent to the ciphertext\\
      $\sum_m\alpha_m|mG\oplus e\rangle$.
  }
\end{itemize}

It is worth to notice that, with respect to Bob, the two states $\sum_m\alpha_m|mG$\\$\oplus e\rangle$ and $X(eG^{-})\sum_m\alpha_m|m\rangle$ are essentially inequivalent. Bob cannot decrypt quantum message from the latter since that there is a trapdoor in the one-way quantum transformation $\sum_m\alpha_m|m\rangle\longrightarrow \sum_m\alpha_m|mG\oplus e\rangle$, but not in $\sum_m\alpha_m|m\rangle\longrightarrow X(eG^{-})\sum_m\alpha_m|m\rangle$. This problem has been thoroughly investigated and developed into a general method for constructing trapdoor one-way transformation in Ref.~\cite{yang2011}. $\square$


\begin{itemize}
  \item {\it The relation of two states with identical probability distribution over some bases.}
\end{itemize}

Conjugate bases measurement on the couple of states $X(eG^{-})\sum\nolimits_m\alpha_m|m\rangle$ and $\sum\nolimits_m\alpha_m|m\rangle$ will result in the identical statistical probability\cite{fujita2012}.
This is obvious since the two states differ only in some bit-flips\cite{nielsen}.
Fujita points out \cite{fujita2012} that the attacker may obtain some information about the quantum message via measurement on $X(eG^{-})\sum\nolimits_m\alpha_m|m\rangle$. His analysis is right though we would like to stress that the similarity of states $X(eG^{-})\sum_m\alpha_m|m\rangle$ and $\sum_m\alpha_m|m\rangle$ is described by the fidelity
\begin{eqnarray}
F(r) &=& \left|\left(\sum_m\alpha_m^*\langle m|\right)X(r)\left(\sum_n\alpha_n|n\rangle\right)\right|\nonumber\\
&=& \left|\sum_{m,n}\alpha_m^*\alpha_n\langle m|X(r)|n\rangle\right|=\left|\sum_m\alpha_m^*\alpha_{m\oplus r}\right|,
\end{eqnarray}
where $r=eG^{-}$ is a random string depending on the error $e$. It can be seen that $F(r)$ may equals to any value from 0 to 1, then, generally speaking, identical probability distributions do not means identical states.
$\square$
\\

In addition, we would like to mention that it is sufficient to adopt once-encryption scheme in some low-level security scenario besides encrypting classical message.
According to Holevo's theorem, the quantum measurement on
$X(eG^{-})\sum_m\alpha_m|m\rangle$ can obtain at most $k$-bit information, but $\sum_m\alpha_m|m\rangle$ has $2^k$ amplitudes $\alpha_m$, and each $\alpha_m$ is $l$-bit complex number which has both real and imaginary parts, so it is necessary to obtain $2l\times2^{k}$-bit information for determining an unknown state $\sum_m\alpha_m|m\rangle$. Even if Alice encrypts the same quantum state polynomial times, the attacker can only obtain at most a polynomial-bits information. We can see that it is still hard for her to determine the state $\sum_m\alpha_m|m\rangle$.

\section*{Acknowledgement}
This work was supported by the National Natural Science Foundation of China under Grant No. 61173157.

\end{document}